\documentclass[pdflatex,sn-mathphys-num]{sn-jnl}

\usepackage{graphicx}%
\usepackage{multirow}%
\usepackage{amsmath,amssymb,amsfonts}%
\usepackage{amsthm}%
\usepackage{mathrsfs}%
\usepackage[title]{appendix}%
\usepackage{xcolor}%
\usepackage{textcomp}%
\usepackage{manyfoot}%
\usepackage{booktabs}%
\usepackage{algorithm}%
\usepackage{algorithmicx}%
\usepackage{algpseudocode}%
\usepackage{listings}%
\usepackage{braket}%
\usepackage{circuitikz}%
\usepackage{cancel}%
\usepackage{comment}%
\usepackage{anyfontsize}

\begin{document}

\title{Selective decoupling in multi-level quantum systems by the SU(2) sign anomaly
.}


\author*[1,3]{\fnm{Giorgio} \sur{Anfuso}}\email{giorgio.anfuso@dfa.unict.it}
\author[2]{\fnm{Giulia} \sur{Piccitto}}\email{giulia.piccitto@unict.it}
\author[2]{\fnm{Vittorio} \sur{Romano}}\email{vittorio.romano@unict.it}
 
\author[1,3,4]{\fnm{Elisabetta} \sur{Paladino}}\email{elisabetta.paladino@dfa.unict.it}
\author[1,3]{\fnm{Giuseppe A.} \sur{Falci}}\email{giuseppe.falci@unict.it}

\affil[1]{\orgdiv{Dipartimento di Fisica e Astronomia ``Ettore Majorana"}, \orgname{Università di Catania}, \orgaddress{\street{Via Santa Sofia 64}, \city{Catania}, \postcode{95123}, \state{Italy}}}

\affil[2]{\orgdiv{Dipartimento di Matematica e Informatica}, \orgname{Università di Catania}, \orgaddress{\street{Viale Andrea Doria 64}, \city{Catania}, \postcode{95123}, \state{Italy}}}

\affil[3]{\orgdiv{Istituto Nazionale di Fisica Nucleare}, \orgname{Sezione di Catania}, \orgaddress{\street{Via Santa Sofia 64}, \city{Catania}, \postcode{95123}, \state{Italia}}}

\affil[4]{\orgdiv{CNR-IMM}, \orgname{Catania (University unit), Consiglio Nazionale delle Ricerche}, \orgaddress{\street{Via Santa Sofia 64}, \city{Catania}, \postcode{95123}, \state{Italia}}}

\abstract{We investigate dynamical decoupling operated by $2\pi$-pulses in a two-level subspace of a multilevel system showing that it may lead to selective decoupling. This provides a flexible strategy for decoupling transitions in a quantum network, when control to directly address them is not available, which can be used to control internode interaction or actively suppress decoherence.
}




\maketitle

\section{Introduction}\label{sec:introduction}

Finding strategies to mitigate the effects of noise is a key issue in quantum technologies. Among these, dynamical decoupling (DD) has attracted a great deal of interest in several platforms~\cite{PaladinoRMP, PRA_Viola1998, vitanov_dynamical_2015,PRAStassiNori2018,DArrigo_2024,Darrigo2014Annals,falci2004PRA,falci2005PhysE}. It is based on the idea that a sequence of pulses can revert the sign of operators in the Hamiltonian that are ``orthogonal" (in the Hilbert-Schmidt sense) to the generators of the pulses. A given component of the Hamiltonian can be effectively attenuated by shining sequences of short equidistant pulses that reverse its sign at alternating intervals.
Complete cancellation is achieved in the limit of an infinitely large pulse rate.  More efficient sequences achieve substantial cancellation at lower pulse rates~\cite{PaladinoRMP} such as the Uhrig sequence~\cite{Uhrig2008}, which reduces decoherence due to low-frequency ``pure dephasing" noise affecting a two-level system.

Even though DD has been widely investigated for two-level systems, the growing interest in the application of qudits to quantum technologies requires the extension of these studies to the domain of multilevel systems.
Extending the geometrical framework, a possible strategy can be identified at least ideally by arguing that interactions that are off-diagonal in the ``logic" basis of the eigenstates of the uncoupled system can be asymptotically mitigated through sequences of $\pi$-pulses with the corresponding diagonal generator.
In particular, this approach could allow for the decoupling of interactions involving specific transitions. To this end, control Hamiltonians oriented in specific directions must be available, which is often not the case in real quantum hardware.

To address this issue, we propose to exploit the $\mathrm{SU}(2)$ sign quantum anomaly by using $2\pi$ pulses on a transition different from the one we wish to decouple. This approach offers two advantages: (i) selectivity, since it cancels only one interaction while preserving the other, and (ii) enhanced control flexibility, since the $2\pi$ pulse can be implemented by any two-level operator acting on the operating transition. Consequently, protocols based on the sign of $\mathrm{SU}(2)$ significantly expand the toolbox for dynamical decoupling in multilevel systems. 

The paper is structured as follows. In Sec.~\ref{sec:model} we introduce the model. The results are discussed in Sec.~\ref{sec:results}. In Sec.~\ref{sec:conclusion} we draw the conclusions.

\section{Model}\label{sec:model}

In a 3-Level System (3LS) with logic basis $\{\ket{g},\ket{e},\ket{f}\}$, we look for a DD protocol that selectively cancels off-diagonal $g-e$ elements, i.e. it yields a dynamics governed by an effective Hamiltonian $H_T$ block diagonal in the subspaces spanned by $\{\ket{g}\}$ and by $\{\ket{e}, \ket{f}\}$.

The dynamics alternates free evolution and an even number $n$ of instantaneous $2\pi$ rotations in the subspace $\mbox{span}\{\ket{e}, \ket{f}\}$ described by the operator $R = \mathrm{e}^{- i \pi \sigma^{ef} }$ where $\sigma^{ef}$ is a $\mathrm{SU}(2)$ generator in that subspace. 
\begin{figure}[!t]
\centering
\begin{circuitikz}
\draw [line width=0.7pt,short        ] (14.5,10.5) -- (10.5,10.5);
\draw [line width=0.7pt,dashed       ] (10.5,10.5) -- ( 9  ,10.5);
\draw [line width=0.7pt,->, >=Stealth] ( 9  ,10.5) -- ( 4.5,10.5);
\draw [line width=0.7pt,short        ] (12.5,10.5) -- (12.5 ,11.5);
\draw [line width=0.7pt,short        ] (10.5,10.5) -- (10.5 ,11.5);
\draw [line width=0.7pt,short        ] ( 9  ,10.5) -- ( 9   ,11.5);
\draw [line width=0.7pt,short        ] (6.75,10.5) -- ( 6.75,11.5);
\node at (12.5,11.75) {$R$};
\node at (10.5,11.75) {$R$};
\node at ( 9  ,11.75) {$R$};
\node at (6.75,11.75) {$R$};

\node at (12.5,10) {$t_1$};
\node at (10.5,10) {$t_2$};
\node at ( 9  ,10) {$t_{n-1}$};
\node at (6.75,10) {$t_n$};

\node at ( 5.8,11) {$\mathrm{e}^{-iH\tau_{n}}$};
\node at ( 8  ,11) {$\mathrm{e}^{-iH\tau_{n-1}}$};
\node at ( 9.8,11) {...};
\node at (11.5,11) {$\mathrm{e}^{-iH\tau_1}$};
\node at (13.5,11) {$\mathrm{e}^{-iH\tau_0}$};
\draw [ color=black , fill=black] (14.5,10.5) circle (1pt);
\draw [ color=black , fill=black] (5,10.5) circle (1pt);

\draw [ color=black , fill=black] (12.5,10.5) circle (1pt);
\draw [ color=black , fill=black] (10.5,10.5) circle (1pt);
\draw [ color=black , fill=black] ( 9  ,10.5) circle (1pt);
\draw [ color=black , fill=black] (6.75,10.5) circle (1pt);

\node at (14.5,10) {$t_0=0$};
\node at (5,10) {$t_{n+1}=T_f$};
\end{circuitikz}
\caption{Diagrammatic representation of a generic sequence of n pulses}\label{fig:pulses}
\end{figure}
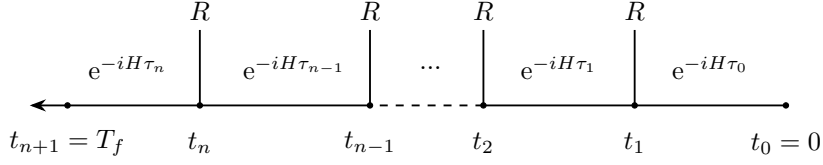
\\
We divide the total evolution time $T_f$ into $n+1$ intervals $I_i\equiv[\,t_i\,;\,  t_{i+1}\,]$ of duration $\tau_i = \delta_i T_f$ where $\delta_i \in ]0, 1[$ for $i = 0, \dots, n$ and
\begin{equation}\label{eq:deltini}
    \ \sum_i \delta_i = 1\ . 
\end{equation}  
The system evolves freely for $\tau_0$, then $R$ is applied and again it evolves freely for $\tau_1$ continuing the pulse sequence until the end (see Fig.~\ref{fig:pulses}). The evolution operator at the final time is
\begin{equation}
    U^n(T_f) =  \mathrm{e}^{-iH\tau_n}  R \, \mathrm{e}^{-i H\tau_{n-1}} R \dots R \, \mathrm{e}^{-i H\tau_{0}}.
    \label{eq:evolution_R}
\end{equation}
Now, the rotation $R$ attaches to the subspace $\mathrm{span}\{\ket{e}, \ket{f}\}$ a phase $\mathrm{e}^{\pm i \pi} = -1$ relative to $\{\ket{g}\}$. 
This results in a dynamical change of sign of off-diagonal $g-e$ elements in the rotated Hamiltonian, defined as  $H_R := RHR$, these being the couplings we would like to suppress. To this end, a pulse sequences must be designed yielding the effective $H_T = \frac{1}{2}(H+H_R)$. 

The success of canceling unwanted couplings depends on the sequence of time intervals $\{\tau_i\}$, that is, the set $\{\delta_i\}$ has to be properly chosen. 
To this end, the effective Hamiltonian $H_\text{eff}(T_f)$, defined by inverting $U(T_f) \equiv \mathrm{e}^{-i T_f H_\text{eff}(T_f)}$, will be studied using analytical approximations obtained from the Magnus expansion. 
For notational simplicity, from now on we will drop the explicit dependence of $H_\text{eff}$ on $T_f$.

The Magnus expansion~\cite{Magnus1954} is a mathematical tool for solving differential equations of the form $Y'=A(t)Y $, being $Y$ a $k-$dimensional vector and $A(t)$ a time-dependent linear operator. This method is useful when $[A(t_1), A(t_2)] \ne 0$, for $t_1, t_2$ belonging to the integration window.
The formal solution reads $Y(t) = T\exp\big(\int_0^t A(s) ds\big) Y_0$, with $Y_0 = Y(t_0)$ being the initial condition, which the Magnus expansion expresses as an asymptotic series $Y(t) = \exp\big(\sum_{k=0}^\infty \Omega_k(t)\big) Y_0$, the first three $\Omega_k(t)$ matrices being~\cite{BLANES2009151}:
\begin{equation}\label{eq:magnus}
    \begin{aligned}
        &\Omega_1(t) = \int_0^t A(t_1) dt_1,  \quad \Omega_2(t) = \frac{1}{2}\int_0^t \int_0^{t_1} [A(t_1), A(t_2)] dt_1 dt_2, \\
        &\Omega_3(t) = \frac{1}{6}\int_0^t \int_0^{t_1} \int_0^{t_2} \Big( [A(t_1),[A(t_2), A(t_3)]+[A(t_3),[A(t_2), A(t_1)]\Big) dt_1 dt_2 dt_3. 
    \end{aligned}
\end{equation}
If $A(t)$ is a piecewise constant matrix, the Magnus expansion can be cast into a generalized Baker-Campbell-Hausdorff formula (see Section 2.8 of ~\cite{BLANES2009151}).

\section{Results and discussion}\label{sec:results}
We start the analysis by rewriting Eq.~\eqref{eq:evolution_R} as
    \begin{equation}
    U^n(T_f) =  \mathrm{e}^{-iH\tau_n} \mathrm{e}^{-i H_R \tau_{n-1}} \dots \mathrm{e}^{-i H_R \tau_{1}} \mathrm{e}^{-i H\tau_{0}} =: T\mathrm{e}^{-i\int_0^{T_f}  \mathcal{H}(t)dt },
\end{equation}
which defines 
\begin{equation}
     \mathcal{H}(t)\Big|_{t \in I_i } = 
    \begin{cases} 
    H \quad \ \ i \ \  \text{is even}; \\
    H_R \quad i \ \  \text{is odd}.
    \end{cases}
\end{equation}
Letting $A(t) = -i \mathcal{H}(t)$ we will use the results of the previous section to evaluate $H_\text{eff}$ at the various orders of the Magnus expansion. Then, order by order, we will look for the best decoupling sequence making $H_\text{eff}$ close to the target $H_T$ at least for the couplings that we want to cancel selectively.
For a $n$-pulse sequence we have to determine $n+1$ time intervals, thus we need $n+1$ equations. For $n=2$ we need 3 equations: one is the constraint Eq.~\eqref{eq:deltini}, 
the others are obtained from the first and the second order of the Magnus expansion, Eq.~\eqref{eq:magnus}. 
The piecewise constant $\mathcal{H}$ yields for the effective Hamiltonian at first order 
\begin{equation}
    H_\text{eff}^1 = \frac{1}{T_f}\int_0^{T_f} \mathcal{H}(t) dt = \Bigg[H\sum_{j \ \text{even}}^n\delta_j + H_R\sum_{j \ \text{odd}}^n \delta_j \Bigg],
\end{equation}
where $H_\text{eff}^1 := \frac{i}{T_f} \Omega_1(T_f)$, which reduces to the target Hamiltonian $H_T$ if
\begin{equation}\label{eq:deltini_1}
    \quad\sum_{j \ \text{even}}^n\delta_j = \sum_{j \ \text{odd}}^n\delta_j =\frac{1}{2}\quad .
\end{equation}
In order to preserve $H_T$ we require that all the following terms of the Magnus expansion vanish.
Looking at second order from Eq.~\eqref{eq:magnus} we can compute
\begin{equation}
    H_\text{eff}^2 = \frac{-i}{T_f}\frac12\int_0^{T_f} \int_0^{t_1}[\mathcal{H}(t_1),\mathcal{H}(t_2)] dt_1dt_2\  ,
\end{equation}
Once again exploiting the structure of $\mathcal{H}$, decomposing the integration into all the possible intervals and observing that
\begin{equation}
     \bigg[\mathcal{H}(t_1),\mathcal{H}(t_2)\bigg]_{^{t_1 \in I_j}_{t_2 \in I_k} } = \left\lbrace
    \begin{array}{cl}

        0 &    \ \ \text{if } \ \ (-1)^j=(-1)^k\ \ ; \\
        (-1)^j[H,H_R] &    \ \ \text{otherwise }\ \ ,
    \end{array} \right.
\end{equation}
after some computation one can find that
\begin{equation}\label{eq:deltini_2}
    H_\text{eff}^{2} = \frac{iT_f}{2}[H,H_R] \sum_{j=1}^{n/2}\left[\delta_{2j-1}\left( \sum_{k=0}^{j-1} \delta_{2l}  \right)-\delta_{2j}\left( \sum_{k=0}^{j-1} \delta_{2l+1}  \right)\right]   ,
\end{equation}
In particular, for $n=2$ and $n=4$ pulses we obtain
\begin{equation}
    H_\text{eff}^2 = \left\{
    \begin{array}{ll} \dfrac{iT_f}{2}\left[H_R,H\right]\delta_1(\delta_0-\delta_2) & \text{if} \quad n = 2\\  \\
        \dfrac{iT_f}{2}\left[H_R,H\right]\left[(\delta_1+\delta_3)(\delta_4-\delta_0)+\delta_2(\delta_1-\delta_3)\right] & \text{if} \quad n = 4\\ 
    \end{array}\right.
\end{equation}
For $n=2$ Eq.~\eqref{eq:deltini}, Eq.~\eqref{eq:deltini_1} and Eq.~\eqref{eq:deltini_2} are the minimal set of equations giving the exact solution,   $\delta_0=\delta_2=1/4$ and $\delta_1=1/2$. 
Instead, for $n>2$ this set of equations is not enough to determine a unique solution, thus we investigate higher order terms. For $n=4$ the third order contribution of Magnus expansion is :
\begin{equation}\label{eq:3ord}
    \begin{array}{rcl}
          H_\text{eff}^3=-iT_f^2\Biggl\{&\left[ H, \left[ H_{R}, H \right] \right]   &\Biggl[-\dfrac{1}{2}\Bigl( \delta_1 + \delta_3\Bigr)\Bigl(\delta_2^2+\delta_4^2+\delta_0^2-4\delta_4\delta_0\Bigr)+\\
          &&\\
          && +\delta_1\delta_2\Bigl(2\delta_0-\delta_4\Bigr)+\delta_2\delta_3\Bigl(2\delta_4-\delta_0\Bigr)\Biggr]\\
          &&\\
          +&\left[ H_{R},   \left[ H_{R}, H \right] \right] & \Biggl[\dfrac{1}{2} \Bigl(\delta_1+\delta_3\Bigr)^2\Bigl(\delta_0+\delta_2+\delta_4\Bigr)-3\delta_1\delta_2\delta_3\Biggr]\Biggr\}
    \end{array}\ .
\end{equation}
We find two more different equations, one for each nested commutator appearing, providing the $n+1$ equations needed to find a unique solution to the optimization problem. However, the resulting system of equations has no real (i.e. physical) solution, thus that the proposed sequences do not yield exactly the target Hamiltonian. 
Still, interesting results can be obtained by relaxing some constraint. In particular, if we only ask the vanishing of the unwanted couplings, the problem admits a class of solutions:
\begin{equation}\left\{
\begin{array}{rl}
   \delta_0&= \frac12\pm\frac{ r(\delta_1)}{8\delta_1}-\frac{\delta_1}{2}, \\
   &\\
   \delta_2&=\frac{1}{8 (1-2 \delta_1) \delta_1\mp 4 r(\delta_1)}, \\
   &\\
   \delta_3&=\frac{1}{2}-\delta_1, \\
   &\\
   \delta_4&=\frac{r(\delta_1)\mp(1-4 \delta_1^2)}{4-8 \delta_1}, \\
\end{array}\right.
\end{equation}
with $r(\delta_1) =\sqrt{16 \delta_1^4-16 \delta_1^3+2\delta_1}$.
By imposing the $\delta_i$ to be real (i.e. the argument of the square root to be positive) and positive we obtain that $\frac{1}{2} - \frac{1}{2 \sqrt{2}} < \delta_1 < \frac{1}{2 \sqrt{2}}$. 
Interestingly, we find that among the possible sequences, the Uhrig one, $\delta_i = \sin^2\Big(\frac{(i+1)\pi}{2n+2} \Big)-\sin^2\Big(\frac{i\pi}{2n+2} \Big)$, is suitable for canceling the 
coupling between the $\{\ket{g}\}$ and $\{\ket{e}, \ket{f}\}$ subspaces.

\section{Conclusion}\label{sec:conclusion}

In this work we have discussed a protocol for selective dynamical decoupling in multilevel systems, by leveraging on the sign anomaly of the $SU(2)$ subalgebra of a $N$-level system.  
We specigfically addressed the case $N=3$, to set the toy model for further numerical analysis, but the idea can be extended to $N>3$ level system problems using the sign anomaly of SU$(M)$ for $2\le M\le N-1$ to average out the leakage from a given subspace of the Hilbert space of a composite system. 

We stress that this approach provides a general framework for multilevel systems that is particularly flexible from the experimental perspective. To investigate the robusteness of the method, we minimised subsequent terms of the Magnus expansion of the evolution operator to find the optimal sequence to cancel the unwanted couplings. The results directly apply to many physical problems, such as selective modulation couplings in quantum architectures in the ultrastrong coupling regime~\cite{falci_advances_2017,falci_ultrastrong_2019,giannelli_detecting_2024},  decoupling and spectroscopy of environmental modes~\cite{bylander2011NatPhys} or quantum sensing\cite{Degen2017} once the correct pulse operators are identified. 

We found that while a sequence that reproduces exactly a given target Hamiltonian does not exist in general, we can engineer sequences giving the desired results up to order $T_f^2$, being $T_f$ the total evolution time. We stress that while sharing analogies with decoupling techniques of "pure dephasing" noise in two-level systems~\cite{PaladinoRMP}, the pulsed control we introduce yields an effective Hamiltonian allowing for decoupling during state processing and tasks of quantum sensing~\cite{DArrigo_2024}. Interestingly, sequences extending Uhrig's one -- very effectively decoupling from pure dephasing noise -- seem to be relevant also for this problem, if shined as $2 \pi$ pulses. 
A natural extension of this work is exploiting sequences with a larger number of pulses and investigating continuous wave control and optimal control methods.

\bf Acknowledgments\rm. 
We thank Antonio D’Arrigo, Matteo Parisi, Nicola Macrì, Giuseppe Chiatto, and Enrico Martello for the fruitful discussions and suggestions.
\backmatter 

\section*{Declarations}

\bf Funding Information\rm. 
GP and GA acknowlwdge support by the
PNRR MUR project PE0000023-NQSTI;
GF and EP acknowledge support from the 
the ICSC – Centro Nazionale di Ricerca in High-Performance Computing, Big Data and Quantum Computing and from 
University of Catania, Piano Incentivi Ricerca di Ateneo 2024-26, project QTCM;
GF is supported from  and the PRIN 2022WKCJRT progect  SuperNISQ;
EP acknowledges the COST Action SUPERQUMAP (CA 21144).

\bf Data Availability Statement\rm. This manuscript has no associated data or the data will not be deposited. [Authors’ comment: All the numerical data in this publication are available on request to the authors.]

\bf Competing Interests\rm. The authors declare they have no financial interests nor competing interest to this work.

\bf Authors contributions\rm. G. Falci proposed and directed the work. G. Anfuso and G. Falci performed the calculation and the numerical analysis, G. Piccitto, E. Paladino e G. Falci supervised the work.



\bibliography{bibliography}

\end{document}